\documentclass[%
reprint,
 amsmath,amssymb,
pra,
]{revtex4-2}

\usepackage{graphicx}% Include figure files
\usepackage{dcolumn}% Align table columns on decimal point
\usepackage{bm}% bold math
\begin{document}

\def \v{\vec}
\def \p{^\prime}
\def \ep{\epsilon}
\def \al{\alpha}
\def \b{\bar}
\def \l{\left}
\def \r{\right}
\def \ba{\begin{array}}
\def \ea{\end{array}}
\def \bea{\begin{eqnarray*}}
\def \eea{\end{eqnarray*}}
\def \f{\frac}
\def \pt{\partial}
\def \bit{\begin{itemize}}
\def \eit{\end{itemize}}
\def \h{\hat}
\def \pp{\perp}
\def \pr{\parallel}
\def \n{\\ \noindent}
\def \th{\theta}
\def \sr{\sqrt}
\def \u{\uparrow}
\def \d{\downarrow}
\def \mf{\mathbf}
\def \be{\begin{equation}}
\def \ee{\end{equation}}
\def \M{{\rm M}}
\def \MAD{{\rm MAD}}
\def \V{${\rm Var}_L$}
\def \E{${\bar E}_L$}

\title{Taming the expressiveness of neural-network wave functions for robust convergence to quantum many-body states
}% Force line breaks with \\

\author{Dezhe Z. Jin}
\email{dzj2@psu.edu}
\affiliation{
Department of Physics and the Center for Theory of Emergent Quantum Matter,  Pennsylvania State University \\
University Park, PA 16802, USA
}

 \homepage{http://www.dezhejinlab.org/}

\date{\today}

\begin{abstract}

Neural networks are emerging as a powerful tool for determining the quantum states of interacting many-body fermionic systems. 
The standard approach optimizes a neural-network ansatz by minimizing the mean local energy estimated from Monte Carlo samples. However, this typically results in large sample-to-sample fluctuations in the estimated mean energy and thus slow convergence of the energy minimization. We propose that minimizing a logarithmically compressed variance of the local energies can dramatically improve convergence. Moreover, this loss function can be adapted to systematically obtain the energy spectrum across multiple runs. We demonstrate these ideas for spin-$\f{1}{2}$ particles in a 2D harmonic trap with attractive Pöschl-Teller interactions between opposite spins.

\end{abstract}

\maketitle

{\it Introduction.--}Variational quantum Monte Carlo (VMC) is a widely used technique for solving many-body wave functions of interacting quantum particles, including electrons in single atoms and molecules \cite{austin2012quantum}, electrons in condensed matter systems \cite{foulkes2001quantum,needs2020variational}, and ultra-cold atomic gases in optical traps \cite{giorgini2008theory}.
The method seeks to optimize a trial wave function $\Psi_\theta(\bm{r})$ in the parameter space $\theta$. Using Monte Carlo algorithms \cite{chib1995understanding}, particle configurations are sampled from the probability distribution defined by the trial wave function, and the local energy $E_L$ is evaluated for each configuration.
The local energy is defined as
$
E_L(\bm{r}; \theta) = H \Psi_\theta/ \Psi_\theta, 
\label{eqn-EL}
$
where $H$ is the Hamiltonian.
The mean $\bar E_L$ of the local energy serves as an estimator of the variational energy associated with the trial wave function. Minimizing $\bar E_L$ can therefore be viewed as optimizing the trial wave function according to the variational principle \cite{foulkes2001quantum}.

For an eigenstate, $E_L$ is constant across all samples. Therefore, the trial wave function can also be optimized by minimizing the variance \V  of the local energy \cite{ceperley1977monte,umrigar1988optimized}. This is known as the zero-variance principle \cite{ceperley1977monte}.
In the early days of VMC, direct minimization of $\bar E_L$ was considered unstable; consequently, minimization of \V was favored \cite{umrigar1988optimized,kent1999monte}. However, it was later shown that the derivative of $\bar E_L$ with respect to the parameters $\theta$ can be written in a form that reduces fluctuations arising from finite sampling, thereby making energy minimization more stable \cite{umrigar2005energy}. This shifted the field away from variance minimization toward energy minimization, which is now the default approach in VMC.

With the rapid advancement of artificial intelligence (AI), neural networks (NNs) have emerged as powerful ansatze for constructing trial wave functions \cite{hermann2020deep,pfau2020ab,von2022self,pfau2024accurate,teng2024solving,qian2025describing}. Their high expressiveness helps overcome the limitations of the hand-crafted trial wave functions traditionally used in VMC. Furthermore, these approaches are naturally well-suited to GPU architectures, enabling efficient and scalable simulations.

In this paper, we show that the expressiveness of NN wave functions can pose a significant challenge for energy-minimization approaches. Specifically, NN wave functions may exhibit a {\it plateau-edge (PE) property} in configuration space, characterized by relatively flat regions connected by sharp edges. In the flat regions, the potential energy can dominate over the kinetic energy, whereas the sharp edges contribute disproportionately to the kinetic energy.
With a finite number of samples, the sharp-edge regions may be missed entirely, yielding an estimate of $\bar E_L$ that is artificially small and can even fall below the true ground-state energy. Conversely, when the edges are sampled, $\bar E_L$ can become large. This leads to large sample-to-sample fluctuations in $\bar E_L$, making energy minimization difficult and highly sensitive to the initialization of the NN parameters. 

We propose minimizing the logarithmically compressed variance as an alternative objective. We show that this loss function enables robust convergence from a wide range of NN initializations, despite the PE property. The energy spectrum can be obtained through multiple runs, with the loss function modified to exclude energies found in previous runs.

\begin{figure*}[t]
	\includegraphics{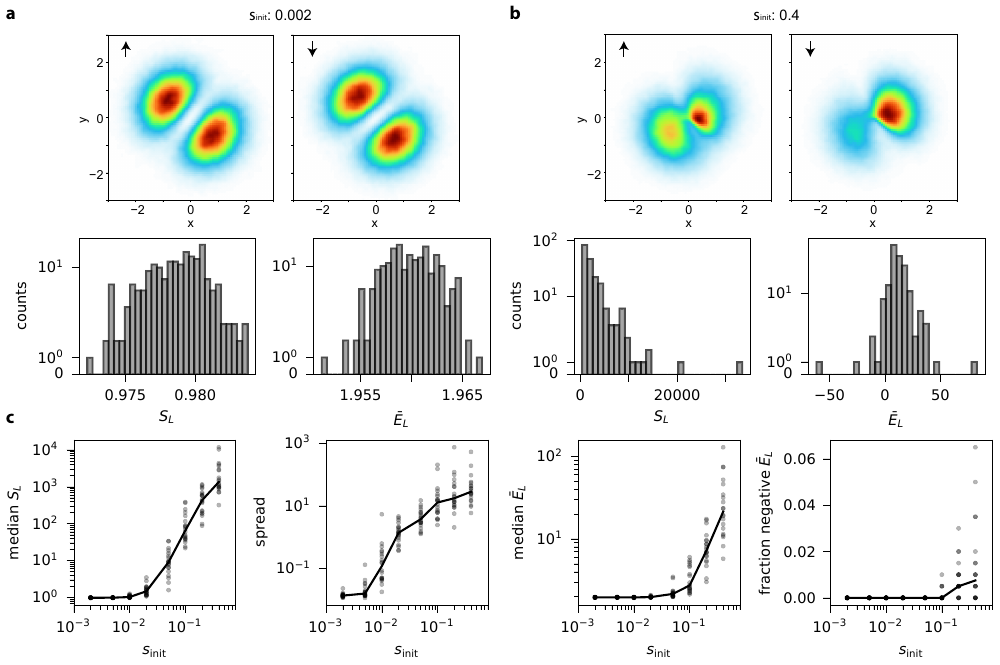}
	\caption{
		{\bf The PE property of NN wave functions}.
		{\bf a}. Spatial distributions of up- and down-spin particles and the distributions of $S_L$ and $\bar E_L$ for $s_{\rm init}$=0.002. The distributions are obtained from 200 sets of 204,800 Monte Carlo samples. The spatial distributions are taken from one representative set.
		{\bf b}. Same as {\bf a} but for $s_{\rm init}$=0.4.
		{\bf c}. Median $S_L$, spread, median $\bar E_L$, and the fraction of sets with negative $\bar E_L$ as a function of $s_{\rm init}$. Each gray dot is obtained from one trial of 200 sets. There are 20 trials, and the thick line connects the median values.
	}
	\label{Fig1}
\end{figure*}

{\it Methods.--}
We study a system of spin-$\tfrac{1}{2}$ fermions confined in a two-dimensional harmonic trap.
Attractive Pöschl–Teller interactions \cite{morris2010ultracold} are present between particles of opposite spin, while particles with the same spin do not interact.
Such systems can be realized in cold atom experiments \cite{holten2021observation}.
The Hamiltonian is given by
$$
H =  \sum_{i=1}^N \frac{1}{2} \left( -\nabla_i^2 + r_i^2 \right)
- \sum_{i=1}^{N_\uparrow} \sum_{j=1}^{N_\downarrow}
\frac{\gamma}{\cosh^2\left( \mu |\mathbf{r}_i - \mathbf{r}_j| \right)} .
$$
Here, $N_\u$ and $N_\d$ denote the numbers of spin-up and spin-down particles, respectively, with $N = N_\uparrow + N_\downarrow$.
The parameters $\gamma >0 $ and $\mu >0$ characterize the strength and range of the Pöschl–Teller potential.
For concreteness, we set $\gamma=1,\; \mu=1$ in this work. 

We use a transformer-based neural network for the trial wave function (Psiformer), consisting of $L$ transformer layers, each with $n_H$ attention heads and embedding dimension $h_H$ \cite{von2022self}. We make two important modifications relative to the original Psiformer: (1) the activation function in the MLP is gelu rather than $\tanh$, and (2) the SoftMax in the attention mechanism is replaced with StableMax \cite{prieto2025grokking}. These modifications improve the stability of the optimization. The output of the final transformer layer is linearly projected to $n_D=40$ sets of $N$ orbitals, which are then used to construct $n_D$ Slater determinants for antisymmetrizing the fermions \cite{lou2024neural}. The final wave function is obtained by summing these Slater determinants and multiplying by the Jastrow factor for the harmonic trap,
$
\mathcal{J}(\bm{r}) = \exp({-\alpha^2 \sum_i^N r_i^2}),
$
where $\alpha$ is a variational parameter.

Particle configurations $\mathbf r$ are drawn from $|\Psi_\theta(\mathbf r)|^2$ using the Metropolis–Hastings algorithm \cite{chib1995understanding}. The NN wave function is optimized using AdamW with weight decay \cite{loshchilov2017decoupled}, a robust optimizer widely used for training neural networks. Each optimization step alternates between computing the gradient of the loss function with respect to the parameters over batches of size 512 and updating the parameters until all samples have been used. A new set of samples is then drawn for the next step.
Optimization is terminated when the standard deviation of the local energy, $S_L={{\rm V}_L}^{-1/2}$, falls below $0.01$. During optimization, the parameters are saved whenever $S_L$ reaches a new minimum. If numerical overflow occurs at any step, the parameters are reverted to the most recently saved values. Optimization is also terminated if $S_L$ does not reach a new minimum for 1000 consecutive steps, or if numerical overflow persists for more than 20 reverts.
The final results are taken from the parameters corresponding to the minimum $S_L$.

The weights of the NN wave functions are initialized with random values drawn from a truncated normal distribution with mean 0 and standard deviation $s_{\rm init}$, discarding values outside $\pm 2\, s_{\rm init} $ \cite{radford2018improving}. Biases are initialized to 0. The Jastrow factor parameter $\alpha$ is initialized to 0.7.

We use the simple case of $N_\u = 1$ and $N_\d = 1$ as a benchmark for investigating the VMC algorithm. The eigenstates can be obtained by separation of variables into center-of-mass and relative coordinates (End Matte (EM)). 

\begin{figure}[h]
	\includegraphics{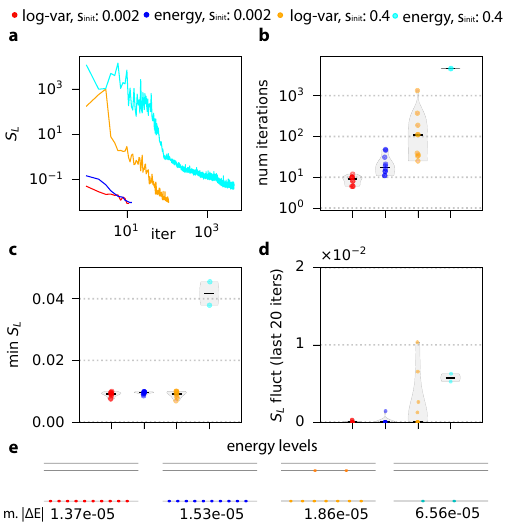}
\caption{
	{\bf Comparison of log-variance minimization and mean-energy minimization for $N_\uparrow = 1$, $N_\downarrow = 1$}. 
	$(L, n_H, h_H) = (2, 2, 2)$. 
	The neural-network wave functions are randomly initialized with $s_{\rm init} = 0.002$ (red, blue) or $s_{\rm init} = 0.4$ (orange, cyan).
	{\bf a}. Examples of $S_L$ as a function of iteration number for log-variance minimization (red, orange) and mean-energy minimization (blue, cyan).
	{\bf b}. Number of iterations required to reach the stopping criterion. Runs for which $S_L$ does not decrease below 0.1 are excluded.
	{\bf c}. Minimum $S_L$ achieved in the runs. 
	{\bf d}. Mean values of $S_L$ over the last 20 iterations before stopping. 
	{\bf e}. Energy levels obtained in the runs. The numbers indicate the median values of $|\Delta E|$ relative to those obtained by exact diagonalization. 
	The black lines in the violin plots in {\bf b}–{\bf d} show the medians.
}
	\label{Fig2}
\end{figure}

{\it Results.--}
We can illustrate the PE property of trial wave functions by constructing simple examples; details are provided in the EM. 
Importantly, the PE property emerges in neural-network wave functions within certain parameter regimes. We demonstrate this for a network with $(L, n_H, h_H) = (2, 2, 2)$.
Initializing neural-network wave functions with small $s_{\rm init}$ tends to produce smooth wave functions, whereas large $s_{\rm init}$ tends to produce more jagged wave functions that can exhibit the PE property (Fig.\ref{Fig1}). We illustrate this for $s_{\rm init}$ ranging from 0.002 to 0.4. For each $s_{\rm init}$, we generate 200 sets of 204{,}800 Monte Carlo samples from the neural-network wave function, yielding distributions of $S_L$ and $\bar E_L$. Figure~\ref{Fig1}a shows these distributions, together with the spatial probability distributions of the up- and down-spin particles. The spatial distributions exhibit broad blobs, suggesting that the neural-network wave function is smooth. Most values of $S_L$ are smaller than 1, and the values of $\bar E_L$ are tightly clustered around 1.96.

\begin{figure}[h]
	\includegraphics{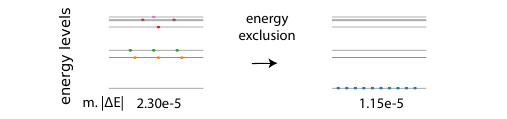}
	\caption{
		{\bf Obtaining the energy spectrum for $N_\uparrow = 1$, $N_\downarrow = 1$}. 
		$(L, n_H, h_H) = (3, 5, 5)$, and $s_{\rm init}= 0.2$. 
		Different runs converge to different energy levels. 
		Excluding energies obtained in previous runs leads to new energy levels. 
	}
	\label{Fig3}
\end{figure}

In contrast, for $s_{\rm init}$=0.4 (Fig.~\ref{Fig1}b), the spatial distributions exhibit finer structures. The values of $S_L$ are broadly distributed with a heavy tail, and $\bar E$ can be negative for some sets of samples. These are hallmarks of the PE property. We quantify the spread of the $S_L$ distribution by
$
{\rm spread}= {(\max S_L - \min S_L)}/{\rm median}\, S_L.
$
For this case, we obtain ${\rm spread} = 18.7$. The median value of $S_L$ is $1.77 \times 10^3$, indicating that the NN wave function is far from an eigenstate.

Increasing $s_{\rm init}$ progressively produces NN wave functions that more strongly display PE property, as shown in Fig.~\ref{Fig1}c. We systematically varied $s_{\rm init}$ from 0.002 to 0.4. For each value of $s_{\rm init}$, we generated 200 sets and repeated this procedure 20 times. We then plotted the median of $S_L$, the spread, the median of $\bar E_L$, and the fraction of sets with negative $\bar E$. All of these measures increase with  $s_{\rm init}$. 

The PE property of NN wave functions can introduce large sample-to-sample fluctuations in \V and $\bar E_L$. This poses a challenge for energy minimization, especially because some samples can have energies lower than the ground-state energy, causing the learning process to drive the NN wave functions further away from the ground state. In contrast, minimizing the log-variance drives the NN wave functions toward eigenstates regardless of these fluctuations  \cite{kent1999techniques,umrigar2007alleviation}. 

We illustrate these points using the case $N_\u = 1$ and $N_\d = 1$, with $(L, n_H, h_H) = (2, 2, 2)$ (Fig.\ref{Fig2}). We compare minimizing the log-variance and the energy. Initialization with $s_{\rm init} = 0.002$ produces smooth NN wave functions (Fig.\ref{Fig1}a). Both log-variance and energy minimization lead to convergence, with $S_L$ reaching 0.01. We compared a set of learning rates and found that $10^{-4}$ and $5 \times 10^{-5}$ work well for minimizing the log-variance and the energy, respectively (EM, Fig.\ref{SFig1}). Ten runs with these learning rates are shown in Fig.\ref{Fig2}. Overall, log-variance minimization reached convergence faster than energy minimization. All runs converged to the ground state.

The situation is markedly different with $s_{\rm init} = 0.4$ initialization, which is expected to produce NN wave functions with the PE property (Fig.~\ref{Fig1}b). For log-variance minimization, 9/10 runs converged to $S_L$ below 0.01, although the numbers of iterations were much larger than in the previous case. One run failed to converge to $S_L$ below 0.1 within 5000 iterations and is excluded. Two of the runs converged to the excited state. In contrast, energy minimization struggled to converge. Only 2/10 runs converged to $S_L$ below 0.1 within 5000 iterations, and none reached below 0.01. These results show that log-variance minimization is much more robust than energy minimization in overcoming the PE property of NN wave functions.

\begin{figure}[h]
	\includegraphics{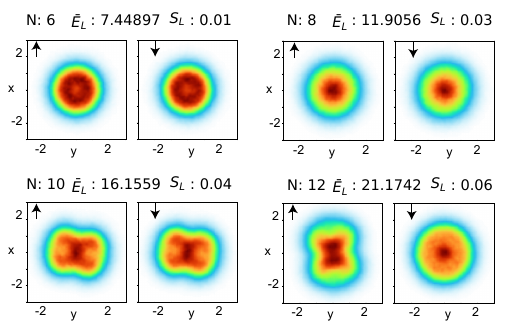}
\caption{
	{\bf Increasing system size}. Here, $N = N_\uparrow + N_\downarrow$ with $N_\uparrow = N_\downarrow$. 
	Particle density, $\bar E_L$, and $S_L$ are shown for $N=6, 8, 10, 12$. 
}
	\label{Fig4}
\end{figure}

We can exploit the robustness of log-variance minimization to obtain the energy spectrum by initializing the NN wave function with a large $s_{\rm init}$. This is demonstrated for the case $N_\uparrow = 1$ and $N_\downarrow = 1$. The network architecture was $(L, n_H, h_H) = (3, 5, 5)$ (Fig.~\ref{Fig3}). The NN wave function was initialized with $s_{\rm init} = 0.2$, and the learning rate was initially set to $10^{-4}$, then reduced to $5 \times 10^{-5}$ once $S_L$ reached 0.05. Across 10 training runs, the learning converged to five distinct energy levels (Fig.~\ref{Fig3}, left). Interestingly, the ground state was not reached.

The search for the energy spectrum can be accelerated by excluding energy levels that have already been obtained. We propose the following modified loss:
$$	\log ({\rm Var}_L) 
-\beta \sum_k  \frac{ {\rm softplus} ({\rm Var}_{exc} - {\rm Var}_{L,k}) }{{\rm Var}_{exc} } \log ({\rm Var}_{L,k}).
$$
Here, ${\rm softplus}(x) = \log (1 + e^x)$ is a smoothed version of the threshold-linear function: it approaches 0 for negative $x$ and approaches $x$ for positive $x$. ${\rm Var}_{L,k}$ is the variance of $E_L - E_k$, where $E_k$ is an energy level that has already been obtained and is therefore to be excluded. The parameter ${\rm Var}_{exc}$ controls the extent of this exclusion. Because of the softplus factor, the exclusion term is effective only when ${\rm Var}_{L,k} < {\rm Var}_{exc}$, that is, when the sampled $E_L$ values are close to $E_k$.

The idea is that if the NN wave function approaches the eigenstate at $E_k$, the loss becomes large, so the optimization dynamics is driven away from this state. The parameter $\beta$ must be large enough to overwhelm the $\log ({\rm Var}_L)$ term. We find that $\beta = 5$ and ${\rm Var}_{exc} = 0.5$ work well. With this modified loss function, all 10 runs converged to the ground state when all five energy levels from the first 10 runs were excluded (Fig.~\ref{Fig3}, right).

Log-variance minimization works well as the system size increases. 
Examples with total particle numbers $N=6, 8, 10, 12$, with $N_\uparrow = N_\downarrow$, are shown in Fig.~\ref{Fig4}. 
As the number of particles increases, the network size must also increase, and more iterations are required to achieve good convergence. 
The network sizes were $(L, n_H, h_H) = (5, 5, 5)$ for the $N=6$ case and $(L, n_H, h_H) = (5, 5, 10)$ for the others. 
The minimum $S_L$ tends to increase with increasing $N$.

{\it Discussion}.--The PE property of neural-network wave functions can be reduced by initializing the network weights with a small standard deviation $s_{\rm init}$.
It is worthwhile to test different values of $s_{\rm init}$, as good initializations often promote faster convergence when training deep neural networks \cite{fort2019goldilocks}.
Another approach is to precondition the neural-network wave function using mean-field reference states such as Hartree--Fock, as is commonly done in VMC simulations of atoms and molecules \cite{hermann2020deep,scherbela2024towards}.
For anti-symmetrization, we used the product of the Slater matrices for the up- and down-spin sectors, as in FermiNet \cite{lou2024neural}.
An alternative is to augment the particle positions with a spin tag and construct a single Slater matrix \cite{von2022self,avdoshkin2025integrated}.
Our preliminary study suggests that this may reduce the PE property of neural-network wave functions and improve convergence for both log-variance and energy minimizations.
For obtaining ground states, combining log-variance minimization with energy minimization remains a viable possibility.

The robustness of log-variance minimization against the PE property allows convergence to different excited states from run to run when the neural-network wave functions are initialized with large $s_{\rm init}$. Using the exclusion method that we propose, different energy levels can be obtained in different runs. This approach to obtaining the energy spectrum is much simpler than existing methods for computing excited states, such as penalizing wave-function overlaps \cite{entwistle2023electronic} or expanding the system size \cite{pfau2024accurate}.

Minimizing the log-variance preserves the gradient information better than minimizing the variance itself
as the variance becomes small. The minimization can be performed using simple first-order optimization techniques such as AdamW \cite{loshchilov2017decoupled}, which is among the most commonly used methods for training AI models, including the large language model GPT-3 \cite{brown2020language}. Compared to second-order optimization methods such as KFAC, which are widely used for optimizing neural-network wave functions \cite{pfau2020ab}, first-order methods are easier to implement and less memory intensive, thereby facilitating scaling to larger system sizes. 

In conclusion, we have shown that log-variance minimization with VMC can control the expressiveness of neural-network wave functions and facilitate the computation of energy spectra in many-body quantum systems.

{\it Acknowledgments.}--This research was supported by the Faculty Upskilling Fellowship in AI and Quantum Sciences, awarded by the Penn State Institute for Computational and Data Sciences. I thank Jainendra Jain, Mytraya Gattu, and Chaoxing Liu for useful discussions.


%%%
% END MATTER
%%%

\clearpage
\newpage
\onecolumngrid

\renewcommand{\thefigure}{S\arabic{figure}}
\setcounter{figure}{0}

\clearpage
\begin{figure*}[t]
	\includegraphics{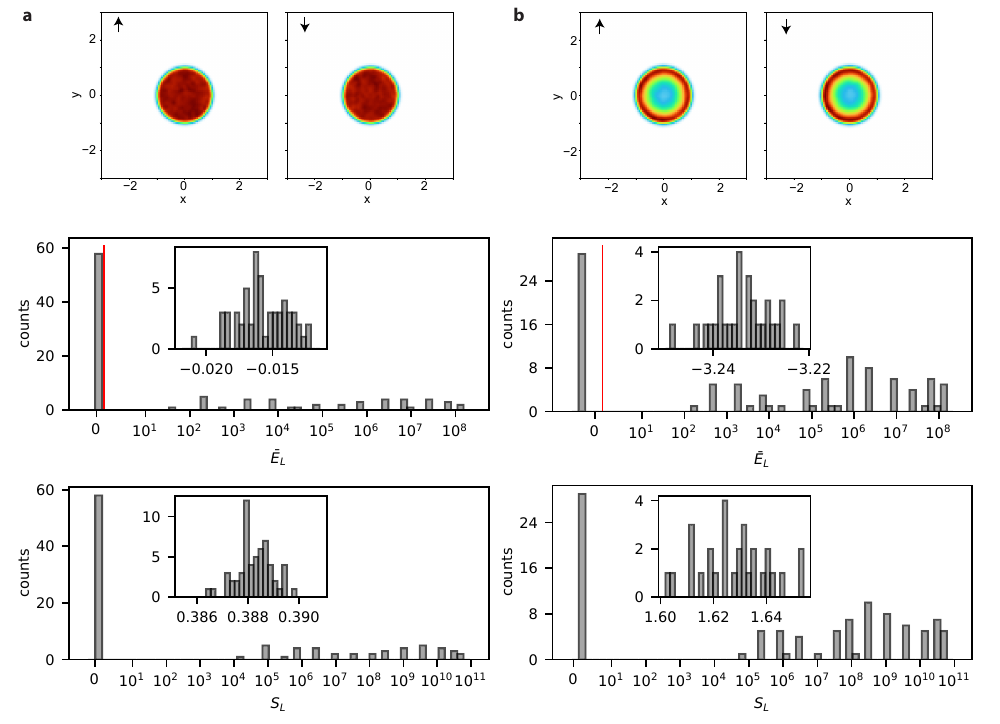}
\caption{Particle densities and distributions of $\bar E_L$ and $S_L$ for 100 sample sets with $N_\u = 1$ and $N_\d = 1$, obtained from two trial wave functions defined by {\bf a} Eq.~\ref{eqn-PE} and {\bf b} Eq.~\ref{eqn-PBE}. The insets show enlarged views of the low-$\bar E_L$ and low-$S_L$ regions. The red line denotes the ground-state energy.}
	\label{SFig1}
\end{figure*}
\twocolumngrid

\section*{End Matter}

{\it Neural network wave function}.--We use a transformer-based neural network for the trial wave function (Psiformer) \cite{von2022self}.
Mathematically, the network can be regarded as a series of transformations of the position vector
$\bm{r}_i$ for each particle, parameterized by the network weights and biases (denoted by $\bm{W}$'s
and $\bm{b}$'s in the following passages).

\begin{table*}[h]
	\begin{tabular}{c | c | c || c | c | c || c | c | c}
		\hline\hline
		(n, m, l) & E & error & (n, m, l) & E & error & (n, m, l) & E & error \\ \hline 
		(0, 0, 0) & 1.592687 & $1.1 \times 10^{-5}$ &
		(1, 0, 0) & 2.592687 & $1.1 \times 10^{-5}$  &
		(0, 1, 0) & 2.830440 & $5.5 \times 10^{-6}$  \\  \hline 
		(2, 0, 0) & 3.592687 & $ 1.1 \times 10^{-5}$ & 
		(0, 0, 1) & 3.800086 & $9.4 \times 10^{-6}$ &
		(1, 1, 0) & 3.830440 & $5.5 \times 10^{-6}$ \\ \hline
		(0, 2, 0) & 3.923556 & $4.8 \times 10^{-6}$ & 
		(3, 0, 0) & 4.592687 & $1.1 \times 10^{-5}$  & 
		(1, 0, 1) & 4.800086 & $9.4 \times 10^{-6}$  \\ \hline 
		\hline 
	\end{tabular}
	\caption{Lowest energy levels for $N_\u =1$, $N_\d=1$.
		$n, m, l$ are quantum numbers. 
		\label{table1}
	}  
\end{table*}

The network consists of $N$ processing streams, one for each particle. In each stream, the particle position $\mathbf r_i$ is linearly projected into a vector $\mathbf R$ in an $M$-dimensional embedding space, which is subsequently transformed by the transformer layers. The stream is influenced by the other particles through the attention mechanism.

Each transformation layer consists of the following sequence:
(1) layer normalization, $\mathbf R_1 = {\rm LayerNorm}(\mathbf R)$, such that the elements of $\mathbf R_1$ have mean $0$ and variance $1$;
(2) addition of an attention vector $\mathbf A$, computed from information from other particles, yielding $\mathbf R_2 = \mathbf R_1 + \mathbf A$;
(3) layer normalization, $\mathbf R_3 = {\rm LayerNorm}(\mathbf R_2)$;
(4) a multilayer perceptron (MLP) with two hidden layers of size $4M$, $\mathbf R_4 = {\rm MLP}(\mathbf R_3)$;
(5) a residual connection producing the layer output, $\mathbf R^\prime = \mathbf R_2 + \mathbf R_4$.
The attention vector $\mathbf A$ is computed using $n_H$ attention heads, where the embedding dimension of each head is $h_H$. We enforce $M = n_H h_H$. The computation follows the standard attention mechanism, which incorporates information from the other particles into the current stream.

{\it Monte Carlo sampling}.--We used 2048 walkers, each generating 100 samples, resulting in 204800 samples overall. Each walker begins from a random configuration and generates a sequence of accepted configurations. To reduce dependence on the initial conditions, the first 1000 configurations are discarded, and to mitigate correlations, every 10th configuration thereafter is recorded as a sample. The acceptance ratio was maintained between $0.2$ and $0.8$.

{\it Optimization method}.—We used the AdamW method \cite{loshchilov2017decoupled} to optimize the NN wave functions. The momentum parameters were set to $\beta_1=0.9$ and $\beta_2=0.999$. The weight decay was set to $10^{-2}$. In simulations with $N>2$, the weight decay was reduced to $10^{-4}$ after $S_L$ reached 0.05. The batch size was 512.

{\it Simple examples of wave functions with PE property}.--Consider the following trial wave function for $N_\u = 1$ and $N_\d = 1$:
\be
\Psi_\theta = \mathcal{N}\, \l [1 + \tanh \l( \f{1 - \rho_1}{d} \r) \r ] \, \l [ 1+ \tanh \l(\f{1 - \rho_2}{d} \r) \r ],
\label{eqn-PE}
\ee
where $\rho_1$ and $\rho_2$ denote the radial positions of the particles, $\mathcal{N}$ is the normalization factor, and $d$ is a parameter. The function defines two plateau regions: for $(\rho_1 < 1, \rho_2 < 1)$ we have $\Psi_\theta \approx 1$, while for $(\rho_1 > 1, \rho_2 > 1)$ we have $\Psi_\theta \approx 0$. Small $d$ creates a sharp crease between the plateau regions.

As a concrete example, we set $d = 10^{-7}$, drew 100 sets of samples from this trial wave function, and computed $\bar E_L$ and $S_L$ for each set. The distributions of $\bar E_L$ and $S_L$ are highly skewed (Fig.~S1a). A fraction of $0.58$ of the sets have $\bar E_L$ smaller than the ground-state energy and exhibit small $S_L$ (inserts). Over all sets, the median value of $\bar E_L$ is $-0.014$, and the median of $S_L$ is $0.39$. Increasing $d$ (reducing the edge sharpness) decreases the fraction of sets with $\bar E_L$ below the ground-state energy, whereas decreasing $d$ (sharpening the edge) increases this fraction.

It is possible to construct trial wave functions that yield negative kinetic energy in the plateau regions. If the curvature of $\Psi_\theta$ has the same sign as $\Psi_\theta$ itself, then the ratio $\nabla^2 \Psi_\theta / \Psi_\theta$ becomes positive, rendering the kinetic-energy contribution negative. In this case, large negative $\bar{E}_L$ values can occur for some samples. One example is
\begin{eqnarray}
	\Psi_\theta &=& \mathcal{N}\, \l [1 + \tanh \l( \f{1 - \rho_1}{d} \r) \r ] \, \l [ 1+ \tanh \l(\f{1 - \rho_2}{d} \r) \r ]  \nonumber  \\
&& \cdot \,(\rho_1^2 + \rho_2^2),
	\label{eqn-PBE}
\end{eqnarray}
which is obtained from Eq.~\ref{eqn-PE} by including an additional factor that introduces positive curvature in the plateau region.

We again set $d = 10^{-7}$ and drew 100 sets of samples. A fraction of $0.29$ of the sets had $\bar{E}_L$ smaller than the ground-state energy, and the energies were more negative than in the previous case (Fig.~\ref{SFig1}b, inset). Increasing the power of the polynomial factor in Eq.~\ref{eqn-PBE} makes $\bar{E}_L$ more negative, while simultaneously decreasing the fraction of sample sets with negative $\bar{E}_L$.

{\it Energy levels for $N_\u=1, \; N_d = 1$}.--We make the coordinate transformation
$
\mathbf r_1 = \mathbf R + \mathbf \rho/2, \; \; 
\mathbf r_2 = \mathbf R - \mathbf \rho/2,
$
where 
$\bm R$ is the center-of-mass coordinates, and $\bm \rho$ is the vector from particle 1 to particle 2. 
Wave function can be separated
$
\Psi(\bm r_1, \bm r_2) = \phi(\bm R) \psi(\bm \rho).
$
The equation for $\bm R$ is that of mass 2 particle in the 2D harmonic trap, 
and the eigenvalues are
$
E_R = n + 1, \; n=0,1,\cdots. 
$ 
Expressing in polar coordinates $\bm \rho = (\rho, \theta)$ and $\psi(\bm \rho) = \eta(\rho) e^{i m \theta}$, 
where $m$ is the angular momentum quantum number, 
we find
\be
\left [ 
- \frac{d^2}{d \rho^2}
- \frac{1}{\rho} \f{d}{d \rho}
+ \frac{m^2}{\rho^2} 
+ \f{1}{4}  \rho^2 
- \frac{1}{\cosh^2(\rho)}
\right ] \eta(\rho) = E_\rho \eta(\rho). 
\ee
The boundary condition is 
$
\eta^\prime (0) = 0, \;
\eta(\infty) = 0.
$

The equation is solved numerically by discretizing the domain $(0, \rho_{\text{max}} = 10.0)$ into $N = 2000$ equally spaced points. The differential operators are approximated using the second-order central-difference scheme, and the resulting discretized equations are formulated as a matrix eigenvalue problem. We systematically varied $m$ from 0 to 10, and for each value of $m$ we computed the lowest 10 eigenvalues corresponding to the quantum number $l$. The lowest energy levels are summarized in Table \ref{table1}. To assess numerical accuracy, the calculation was repeated with halfed grid resolution ($N=1000$), and the absolute differences between the two sets of results were taken as estimates of the numerical error.

\begin{figure}[b]
	\includegraphics{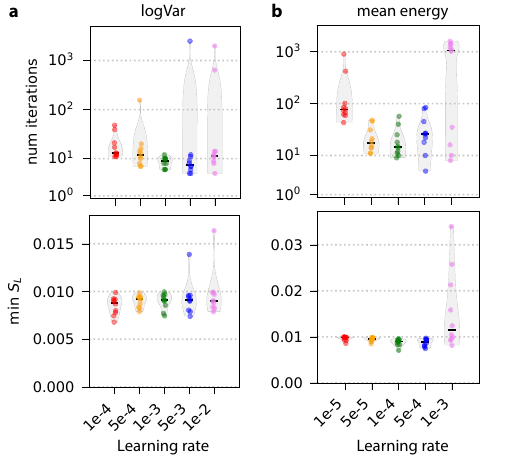}
\caption{
	{\bf Optimization dynamics vs.\ the learning rate for $N_\u = 1$, $N_\d = 1$, and $s_{\rm init} = 0.002$}. 
	$(L, n_H, h_H) = (2, 2, 2)$. There are 10 runs for each learning rate. The number of iterations required to reach the stopping criterion and the minimum $S_L$ attained are shown as violin plots for (a) minimizing the log-variance and (b) minimizing the mean energy.
	The black lines in the violin plots show the medians.
}	
\label{SFig2}
\end{figure}

\end{document}